# Half-Metallicity in Smallest Cage-like Cluster of CdTe with Doping of Transition Metal Atoms

**Kashinath Tukaram Chavan[a], Sharat Chandra\*,[a] and Anjali Kshirsagar[b]**

[a]*Materials Physics Division, Materials Science Group, Indira Gandhi Centre for Atomic Research, HBNI, Kalpakkam 603 102, INDIA*

[b]*Department of Physics, Savitribai Phule Pune University, Pune 411 007, INDIA*

*\*Corresponding author email: sharat@igcar.gov.in*

**Abstract:** We report first principles theory based electronic structure studies of a semiconducting stoichiometric cage-like $Cd_9Te_9$ cluster. Substantial changes are observed in the electronic structure of the cluster on passivation with fictitious hydrogen atoms, in particular, widening of the energy gap between highest occupied molecular orbital and lowest unoccupied molecular orbital and enhancement in stability of cluster is seen. The cluster, when substitutionally mono-doped for a Cd by a set of 3d and 4d transition metal atoms (Ti, V, Cr, Mn, Fe, Co, Ni, Cu, Zn, Ru, Rh and Pd), is found to acquire polarization as seen from spin resolved density of states near Fermi level. Further, such mono-doping in passivated cluster shows half-metallic behavior. Mapping of partial density of states of each system on that of undoped cluster reveals additional levels caused by doping each TM atom separately. In the 3d elemental doping, Ti and Mn doping result into electron type doping whereas all other cases result into hole doped systems. For all the 4d elements studied, it is akin to the doping with holes for Cd substitution in the outer ring, whereas for Ru and Rh, there is electron type doping in case of substitution for Cd in central ring upon passivation. A comparison of partial density of states plots for bare and passivated clusters, on doping with transition metal atoms, suggests suitability of the cage-like cluster for spintronics applications.

## Introduction

The zero-dimensional objects, like an atomic cluster, can provide profound fundamental understanding of quantum mechanical behavior of a system. The clusters have features like high surface to volume ratio, quantum confinement, along with changes in the types of bonding involved. The properties of clusters are basically unique and thus are interesting to understand and exploit. In semiconducting clusters, along with the optical properties, magnetic behavior can also be studied by doping with a magnetic impurity. Clusters of few atoms can be suitable candidates for nanocontacts in electronic devices and thus understanding the phenomenon of electrical transport through small clusters can be complementary to the more involved research of molecular devices. The low-dimensional forms of semiconductors are already being used in various applications. The nanocrystals of semiconductors are being used as fluorescent biological labels [**1-2**] which use their luminescence property. Clusters are also reported to act as super-atoms or building blocks for self-assembled structures [**3**].

The class of compound semiconductors, especially II-VI and III-V, has continued to acquire attention of the scientific community due to their versatility and for potential applications. These classes of materials are also being extensively studied in various forms due to their growing importance in photo-voltaic solar cells **[4-5]**, spintronics **[6-7]**, radiation detectors **[8-9]**, quantum devices **[10]**, opto-electronic devices **[11-12]**, etc.

Cadmium telluride (CdTe) is a II-VI compound semiconductor and has zinc blende (ZB) crystal structure with a direct band gap of 1.5 eV at room temperature in bulk form [**13**]. CdTe exhibits some unique features in the II-VI family, viz. amphoteric semiconducting behavior, highest ionicity, largest lattice parameter and highest average atomic number **[14].** Apart from the solar cell applications, CdTe is an important material for spintronic and radiation detector applications **[15-17].**

Clusters of II-VI semiconductors have been studied for many years. Early theoretical work on structure and stability of II-VI stoichiometric clusters ($Cd_NX_N$; X = S, Se, Te; N = number of formula units) by Matxain et. al [**18**] suggests that the structurally favorable form up to a certain size is the ring and beyond that is the spheroid, basically a cage-like structure. As far as the relative stability of the clusters is concerned, the stoichiometric clusters such as $Cd_6Te_6$, $Cd_9Te_9$, $Cd_{12}Te_{12}$, $Cd_{15}Te_{15}$, etc. in the small to medium size regime of clusters, have been reported to be stable theoretically **[19-20**].

It has been known that bulk CdTe becomes half-metallic when doped with certain 3d and 4d transition metal (TM) atoms [**21-23**] and hence upon doping it can be the source of spin polarized current. The efficiency of spintronics devices depends on the magnitude of polarization in current source. Thus, the half-metals are ideal for spintronics applications. Therefore, the study of such clusters with TM doping can be useful in spintronics applications and detectors. To the best of our knowledge, there are no such reports on whether the half-metallic behavior of CdTe is retained in low dimensions (clusters) when doped with the transition metal (TM) atoms. TM doping in such cage- like clusters can be done in three different ways, namely, endohedral, exohedral and by substitution. Out of these three, substitutional doping is favorable as has been shown in other II-VI clusters [**24-25**].

In this work, we present the results of our first principles, spin-polarized electronic structure calculations to determine the properties of $Cd_9Te_9$ cluster, without and with passivation using fictitious hydrogen-like pseudo atoms (FHPA). The structures are substitutionally mono-doped for a Cd by a set of transition metal atoms viz. Ti, V, Cr, Mn, Fe, Co, Ni, Cu, Zn, Rh, Ru, Pd, at two distinct positions of Cd, independently. We show that the passivated doped clusters are half-metallic in nature and discuss their properties.

**Methodology**

The geometry optimization and further electronic structure calculations are performed using ab initio technique based on density functional theory as implemented in the VASP package **[26-27]**. Projected augmented wave (PAW) pseudopotentials are used in the Perdew–Burke–Ernzerhof (PBE) **[28]** flavour of the generalized gradient approximation (GGA) for the exchange-correlation energy functional [**29**]. The cluster is embedded in a cubic cell with the optimized dimensions of 30×30×30 Å$^3$, chosen to minimize the inter-cluster interactions arising from artificial periodic boundary conditions. The total energy calculations for such zero-dimensional systems are performed using only the Γ point, the center of the Brillouin zone. The criterion used for total energy convergence for self consistency is $10^{-7}$ eV. For all the calculations, the optimized kinetic energy cut-off for plane waves, which determines the size of the basis set, is 700 eV. All the structures are relaxed for a maximum inter-atomic force tolerance of $10^{-3}$ eV/Å.

In order to obtain the ground state configuration of the $Cd_9Te_9$ cluster, we have carried out geometry optimization performed using *ab-initio* molecular dynamics (AIMD). The following procedure is used. A fragment of bulk CdTe is chosen as the initial input. Then this fragment ($Cd_9Te_9$) in the initial ZB structure is made to undergo simulated annealing. Temperature of the system is raised from 0 K to 1400 K, which is near to the melting point of bulk CdTe. Thereafter, constant temperature AIMD simulations are performed for a sufficiently long time (0.2 ns in steps of 1.0 fs) to ensure that the system explored almost the whole potential energy surface. A number of random configurations (500 random structures) are chosen from the production runs. Each of these structures is then statically relaxed to the nearest local potential energy minimum using the conjugate gradient method [**30**]. The lowest energy configuration thus found is taken to be the global minimum and will most probably be the ground state geometry.

The binding energy (BE) per atom is calculated using following formula

$$BE = \frac{[(N \times E_{Cd} + M \times E_{Te}) - E_{Cd_N Te_M}]}{(N+M)}$$

where $E_{Cd}$ and $E_{Te}$ are the total energies of isolated Cd and Te atoms respectively and $E_{Cd_N Te_M}$ is the total energy of the cluster.

ZB structure has tetrahedrally bonded atoms and hence the valences of constituent atoms are satisfied. In the cluster, the bonds on the surface become the dangling bonds. Such bonds can lead to presence of defect states in the energy gap region near the Fermi level which ultimately results into false estimation of the gap between the highest occupied molecular orbital (HOMO) and the lowest unoccupied molecular orbital (LUMO). One can get rid of this artefact and can enhance the stability of cluster by satisfying those dangling bonds by passivating the structure with appropriate ligand(s). Such procedures also make it possible to compare the simulation results with experiments that use various ligands to suppress the growth of the cluster by satisfying the dangling bonds. Though, surface passivation is not always necessary. Puzder et al [**31**] have reported that the relaxation of CdSe nanostructures opens the HOMO-LUMO gap and it does not need to take the aid of a passivating agent to take care of its surface states. This phenomenon of natural reconstruction of surface states is termed as "self-healing". However, this mechanism of self-healing depends on the stoichiometry and constituents of the clusters. Bhattacharya et al. [**19**] have reported that the self-healing mechanism is inefficient in taking care of surface states for CdTe clusters. It is observed that the states in the HOMO-LUMO gap disappear upon surface passivation which results into further widening of energy gap. Therefore, CdTe clusters need to be passivated and it can be done computationally using fictitious hydrogen-like pseudo atoms (FHPA) in order to obtain neutral cluster configurations.

The FHPA (with fractional atomic numbers but otherwise neutral atoms) are chosen as passivating ligands and their Z value is taken such that they satisfy the valence of host atoms

[19]. For Cd and Te atoms, the orbitals $4d^{10}5s^2$ and $5s^25p^4$ respectively, are considered as the valence orbitals for generating the pseudo-potentials. As there is tetrahedral arrangement of atoms in ZB structure, each of the Cd and Te covalently bonded atoms shares 0.5 and 1.5 electron charge with four neighboring Te and Cd atoms, respectively. Thus, the suitable FHPA's Z values for Cd and Te are 1.5 (H*) and 0.5 (H**), respectively. The bond lengths of Cd-H* and Te-H** are optimized and are found to be 1.82 Å and 1.78 Å, respectively and are used as initial bond distances to passivate the clusters. In both the passivated and unpassivated $Cd_9Te_9$, the TM (Ti, V, Cr, Mn, Fe, Co, Ni, Cu, Zn, Rh, Ru, Pd) atom is initially substituted for a Cd on the $Cd_9Te_9$ cage and then the whole structure is relaxed freely till the inter atomic forces are less than the set force tolerance. These relaxed structures are then used for performing further electronic structure calculations.

**Results and Discussions**

As mentioned earlier, sufficiently large number (more than 500) of random geometry configurations, picked from the AIMD production runs, are statically relaxed to the nearest local minimum of the potential energy surface. The one which is found to be the lowest in energy may correspond to the ground state configuration of the cluster. The optimized geometries of the bare and FHPA passivated cluster are shown in figure 1 in different views. This cluster has three-fold rotational symmetry (point group $C_{3h}$) which it retains on passivation also. All the atoms together form a cage, the interior of cluster being empty, corresponding to the spheroid form. $Cd_9Te_9$ cluster has been reported by many researchers and is one of the most stable and the smallest cage-like structure **[18-20].** Our geometry agrees with the previous reports. If at all the cluster dissociates, the most probable fragments are three $Cd_3Te_3$ clusters on supply of energy equal to 3.9 eV. These cluster structures can be viewed in different ways as if comprised of further smaller units. For bare cluster firstly, (parallel to plane of paper) it can be seen to be comprised of three $Cd_3Te_3$ buckled rings: a central ring which connects in parallel to other two identical rings (ABA stacking), or secondly, (perpendicular to plane of paper) three chairs (starting from the outermost Te atom) of $Cd_3Te_3$ interconnected in circular fashion (CCC) and thirdly, as three close connected units of $Cd_3Te_3$ in series. This cluster has five hexagonal and six rectangular faces. This geometry of $Cd_9Te_9$ is consistent with the earlier reports for $Cd_9Te_9$ and $Cd_9S_9$ clusters [**2, 20, 32**]. The rings are buckled due to the Coulomb repulsion between the Te atoms in identical rings as they have an extra lone pair of electrons [19]. As a result of passivation, the three buckled rings of $Cd_3Te_3$ become planer rings, the chairs become boats and both Cd and Te atoms are found on the surface. This reduces the strain in the structure and improves the stability of cluster. The FHPAs point outward from the cage. Upon relaxation, the Cd-H and Te-H bond-lengths change to 1.84 Å and 1.79 Å, respectively. The increased stability of structure upon passivation reflects in the binding energy value as it increases from 1.78 eV per atom in bare cluster to 2.00 eV per atom upon passivation. The average Cd-Te bond lengths of the adjacent sides of the hexagonal faces are 2.88 Å and 2.76 Å in bare cluster which change on passivation to 2.88 Å and 2.82 Å, respectively. Similarly, the average Cd-Te bond length of the surface atoms increases from 2.88 Å to 2.90 Å. In order to know the cluster's thermal stability at room temperature, ab-initio molecular dynamics within the canonical ensemble using Nosè-Hoover thermostat is performed for bare as well as FHPA passivated cluster. Figure 2 shows the fluctuations in total energy for both the forms of cluster as a function of time at 300 K. The total energies for bare and passivated $Cd_9Te_9$ cluster are -38.8687 eV and -83.3892 eV respectively at 0K. The effect of temperature on the total energy seems to be more in the passivated cluster. This may be due to the fact that the attached FHPAs are relatively free to move.

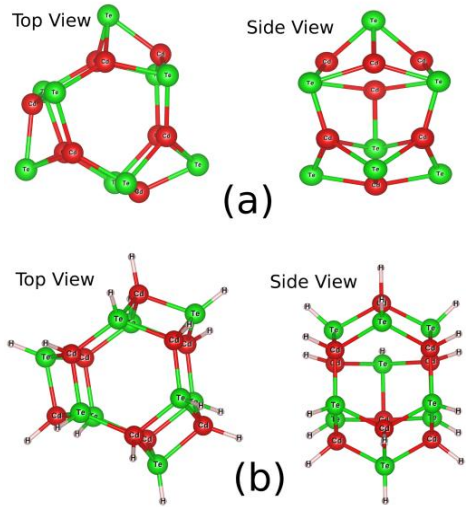

Figure 1. Geometries of (a) bare and (b) FHPA passivated $Cd_9Te_9$ clusters in top and side view. The coloured spheres represents Cd (Red), Te (Green) and FHPA's (Grey) atoms.

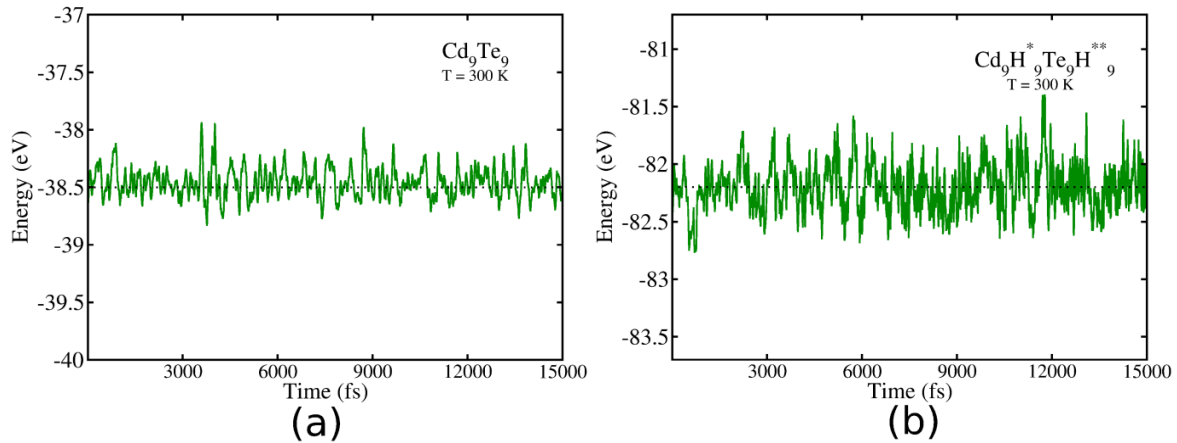

Figure 2. Plots of total energy as a function of simulation time from molecular dynamics at 300K for (a) bare and (b) FHPA passivated $Cd_9Te_9$ cluster.

The *l*-decomposed and atom projected partial density of states (PDOS) of $Cd_9Te_9$ are plotted in figure 3. The states which lie deep below the Fermi level around -8 eV and -10 eV have predominant contribution from Cd-d and Te-s states respectively. The states near Fermi level are mainly populated by Te-p states with minor contributions from the Cd-s and d states and Te-s states. Partially filled 5p states of Te populate the HOMO as well as LUMO of $Cd_9Te_9$. A widening of HOMO-LUMO gap is observed from 1.6 eV to 2.5 eV on passivation. The charge densities corresponding to HOMO (left) and LUMO (right) are also shown at the bottom inset. Analogous to *l*-decomposed PDOS, the HOMO and LUMO charge density in bare as well as passivated cluster also demonstrates the angular character of Te-p states. The top-right inset in figure 3 is the DOS in the energy range -4 eV to 4 eV about the Fermi level. The HOMO and LUMO charge densities reside on the surface, as expected, in the bare cluster due to the presence of dangling bonds with HOMO on Te atoms and LUMO on Cd atoms. Passivation partially satisfies the dangling bonds as seen from the isosurfaces.

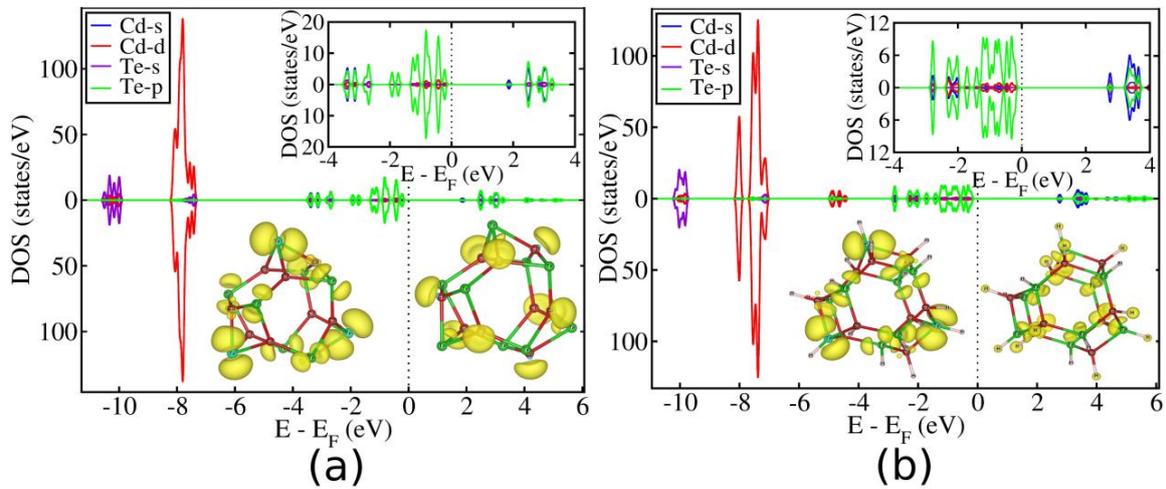

Figure 3. Site projected density of states plots for (a) bare and (b) FHPA passivated $Cd_9Te_9$ clusters. Fermi level ($E_F$), denoted by the vertical dotted line, has been shifted to 0 eV. Inset at top-right shows the DOS in the energy range -4 eV to 4 eV, whereas the bottom inset in each figure denotes the electronic charge density corresponding to HOMO (left of $E_F$) and LUMO (right of $E_F$) for the two cases. These charge densities are plotted at a common isovalue of 0.002 at which the orbital characters are visible. This isovalue ranges from 10% to 20% of maximum value of partial charge density in individual case.

The importance of passivation in the CdTe clusters can be understood through the observed significant changes in the electronic structure upon passivation. The first important but expected result is widening of the HOMO-LUMO gap by segregating the Cd-s and d states and Te-p states and pushing these states further down in energy (it may be noted that the x-axis of PDOS plots is E-$E_F$). Therefore, unlike CdSe, the self-healing mechanism alone is not sufficient to take care of surface states for CdTe and therefore it needs to be passivated **[19]**. Another expected observation is that the LUMO is largely localized on the FHPAs rather than on the Cd and/or Te atoms.

A Cd atom in the bare as well as passivated clusters is substituted by a TM atom either in the central ring (CR) or in the outer ring (OR) of $Cd_9Te_9$ clusters. All the doped TM atoms have either smaller or similar covalent radii as Cd [**33**], therefore even upon doping, the cluster retains its geometrical cage-like spheroid form. (Please refer to figure S1 in Supplementary Information (SI) for all the geometries). Binding energy of the doped cluster(s) is more than the binding energy of the pristine cluster, for both bare and passivated clusters. Also the CR doped cluster is more stable than the respective OR doped ones. TM atom in the cluster does not have any ligands attached; it is directly bonded with three nearest Te atoms. Further, when the TM atom is substituted at the Cd site in the $Cd_9Te_9$ cluster, the states that lie 8 eV to 10 eV below the Fermi level (Cd-d and Te-s i.e., the core states) remain least affected but the valence and conduction states in the ±4 eV energy range about the Fermi level change significantly. Therefore, in the discussion of the PDOS of TM doped cases, we focus only on the energy region near the Fermi level. It may however be mentioned that in case of passivated clusters with doping of Cr and TM atoms with more than half-filled d states, the Cd-d states do move up higher in energy. This may be a result of d-d interaction.

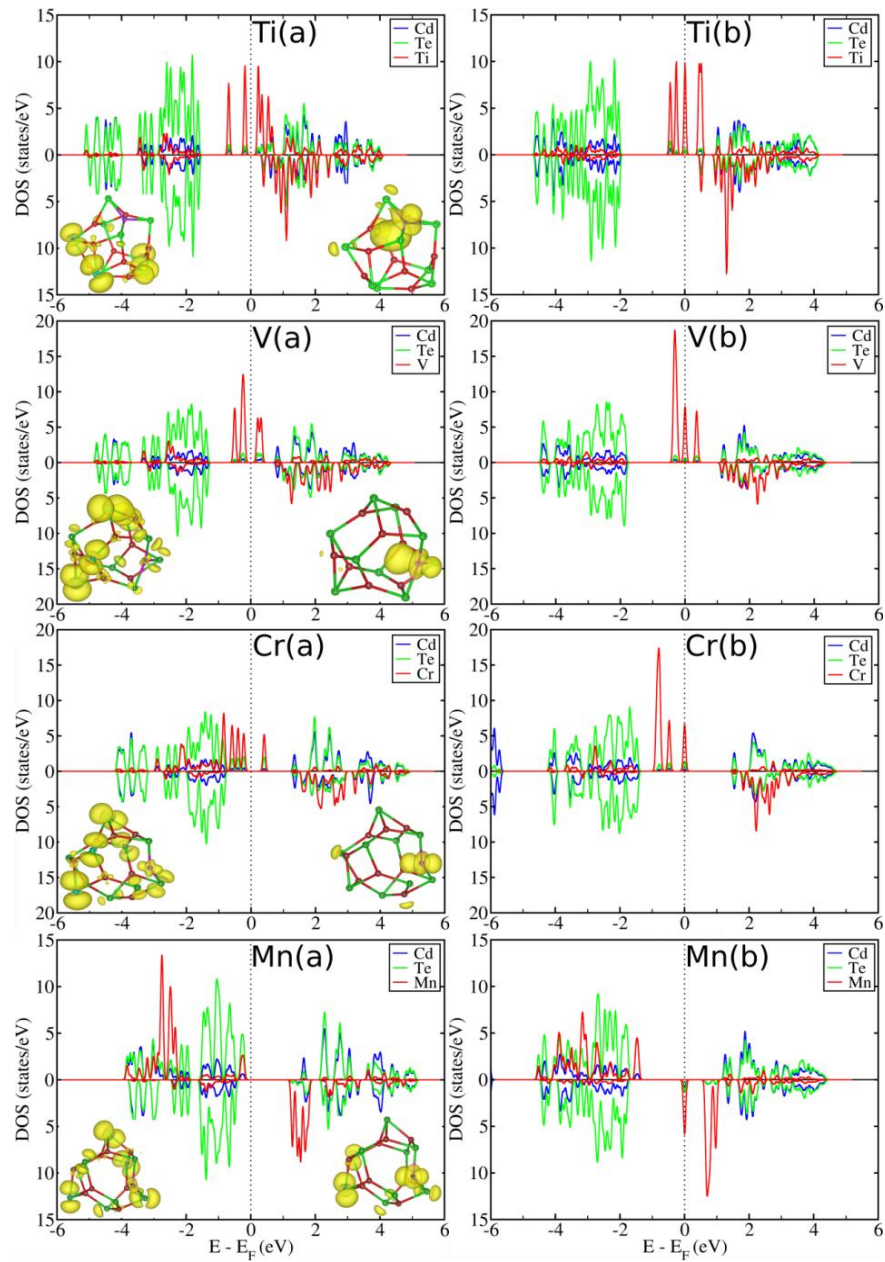

Figure 4. Site projected density of states for TM atom doped (a) bare and (b) FHPA passivated $Cd_9Te_9$ clusters for different 3d TM (Ti, V, Cr Mn) atom substitution for Cd in the OR. Fermi level ($E_F$), denoted by the vertical dotted line, has been shifted to 0 eV. The bare doped clusters are semiconducting in nature, their electronic charge densities corresponding to HOMO (left of $E_F$) and LUMO (right of $E_F$) are shown as inset in corresponding figures.

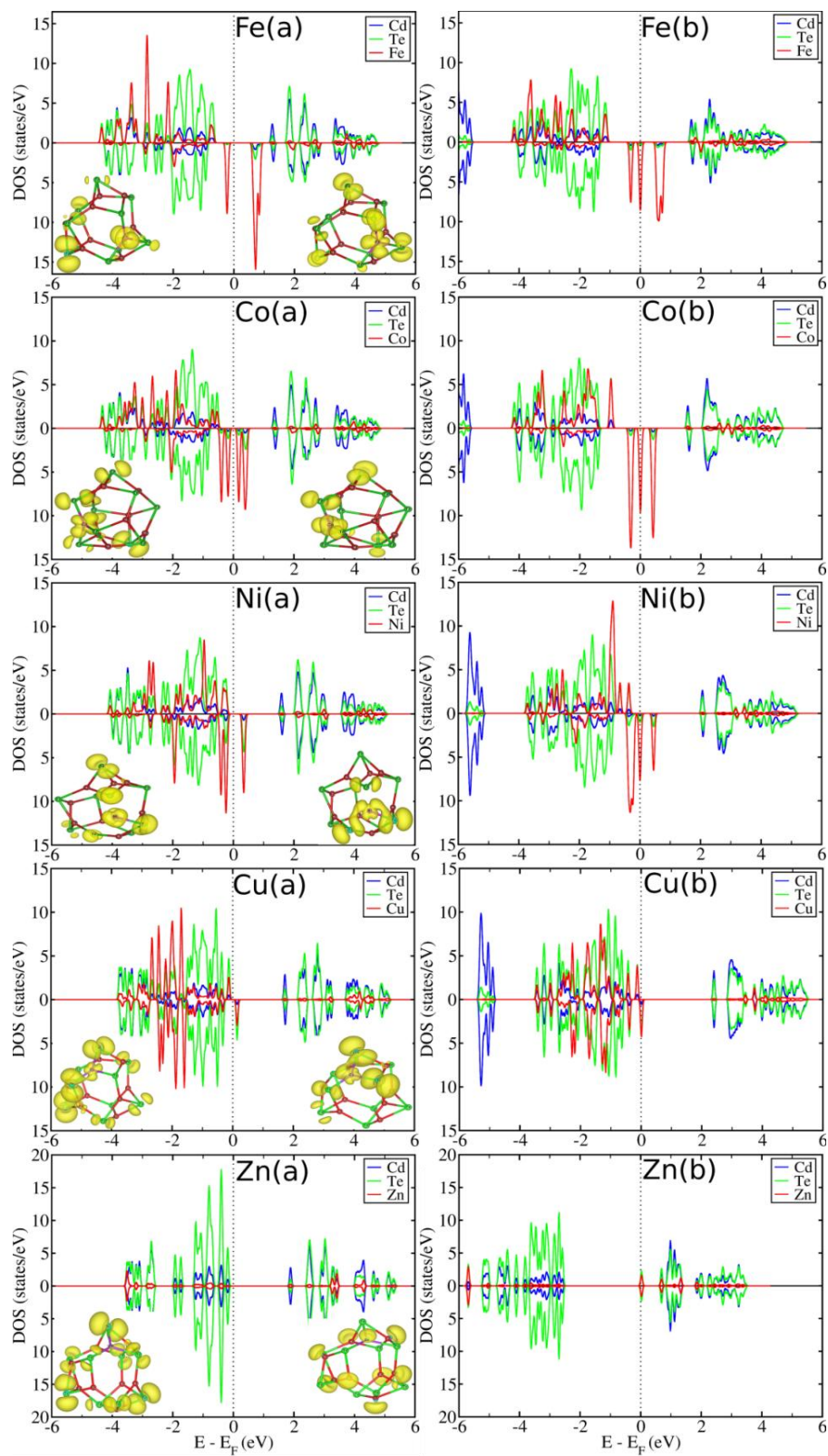

Figure 5. Site projected density of states for TM atom doped (a) bare and (b) FHPA passivated $Cd_9Te_9$ clusters for different 3d TM (Fe, Co, Ni Cu, Zn) atom substitution for Cd in the OR. Fermi level ($E_F$), denoted by the vertical dotted line, has been shifted to 0 eV. The bare doped clusters are semiconducting in nature, their electronic charge densities corresponding to HOMO (left of $E_F$) and LUMO (right of $E_F$) are shown as inset in corresponding figures.

PDOS plots for the 3d TM substituted for a Cd in the OR (CR) of the cluster are shown in figures 4 and 5 (figures S2 and S3 in SI). The PDOS plots labeled TM(a) are for bare substituted cluster and those labeled TM(b) are for the passivated substituted cluster. The nature of PDOS is almost the same on the energy scale for TM substituted in OR and CR with the actual values of PDOS differing slightly for both the bare and passivated clusters except for Cr, Ni and Zn. For semiconducting clusters, the partial charge densities corresponding to HOMO (left) and LUMO (right) are presented in the insets in each plot. Fermi level has been shifted to 0 eV.

The doped 3d TM atoms have d states which are less than half filled, exactly half-filled and more than half-filled for different elements. The HOMO and LUMO of bare cluster are mainly populated by Te-p states with small contribution from Cd-s state. The 3d states of the doped TM atom mainly appear in the HOMO-LUMO gap region of undoped cluster. They also partially hybridize with the Te-p and Cd-s states. The cluster without passivation retains its semiconducting nature with the TM-d states appearing in the HOMO-LUMO gap upon TM substitution. On passivation, we observe half-metallic nature (exception being Zn; it acquires metallic nature) of the DOS in the clusters where it is metal-like for one spin and semiconductor-like for the other spin. As expected, for substitution with TM atom with less than half filled d states, the up or majority spin TM-d states lie in the HOMO-LUMO gap and the down or minority spin TM-d states lie in the unoccupied region while for substitution with TM atom with exactly half-filled or more than half filled d states, the up or majority spin TM-d states lie deep in the valence band region and the down or minority spin TM-d states lie in the HOMO-LUMO gap region The half-metallic nature of the cluster is observed independent of position of the TM substitution, i.e., whether the TM is substituted for Cd in the OR or in the CR. The calculated HOMO-LUMO gaps for each type of carriers in all the cases are listed in table ST2 in the SI.

TM doping in passivated cluster, in general, exhibits less hybridization with the host states as compared to bare doped cluster and the TM d states are more localized. For Ti, V, Cr and Mn substitution, hybridization of Te-p and TM-$d_{z^2}$ is seen in the charge density plots. Both up and down Ti-d states are near the LUMO. The up spin V-d states are midway in the HOMO-LUMO gap and down spin d states merge in the LUMO. Substitution of Cr with exactly half-filled 4s and 3d states, according to Hund's rule, need special mention because of its unique electronic configuration. The up and down spin Cr-d states are fairly separated and therefore Cr doped cluster is a good candidate for spintronic applications. Also due to the unique electronic configuration, position of Cr atom (CR or OR) affects the electronic structure. Mn atom with completely filled 4s and exactly half-filled 3d states also has unique ordering of states in energy; 3d up states are lowest in energy followed by 4s up, 4s down states and all are occupied. The 3d down states are unoccupied. Therefore the up and down states are well separated with up states hybridizing with the Te-p states and lying deep in the valence band region while the down states lie just below the LUMO of host consisting of Te-p and Cd-s states. For all the TM atoms with more than half filled 3d states, the up spin states are deep in the valence region and are completely occupied while the down spin states lie in the HOMO-LUMO gap and are partially occupied. Ni and Cu have even their down states close to HOMO and partially occupied. Cd-d states in these clusters have moved to higher energies.

When the TM atom with completely filled d and s states, Zn, is doped in the bare cluster at any position (OR or CR), as expected, it does not lead to any polarization in PDOS. The states near the Fermi level have contribution from Te-p and Cd-s states and there are no states due to Zn atom. On passivation there is shift of states to lower energy with widened gap. In OR case the states due to both the carriers appear at Fermi level and system acquires metallic nature. The states at Fermi level have comparable contribution from Te and Zn atoms. For CR substitution, there is little polarization in PDOS. States exist at the Fermi level due to majority charge carriers of Zn-s and Cd-s leading to half-metallic nature of system.

The PDOS of all the doped systems with and without passivation are plotted without shifting the Fermi level by mapping over the PDOS of corresponding host system (bare or passivated cluster) for comparison and are presented in figures 6 and 7 for CR substitution. (Please refer to figures S4 and S5 for OR substitution.) The region between the two vertical dotted lines denotes the energy gap region of the host $Cd_9Te_9$ cluster. We now see that the changes in the properties of the cluster are mainly due to the defect states that are contributed by the different dopants in energy gap of the undoped cluster. In FHPA passivated cluster case, the Fermi level gets pinned in one of the defect maximum in the PDOS and leads to the observed half-metallic behaviour.

Ti atom doping introduces states in the middle of the energy gap of the cluster (deep states) and donor levels near the LUMO in all the four cases viz., CR and OR substitution with and without passivation. V atom doping too introduces deep defect levels in all four cases. Te contribution to those deep states changes from small to vanishingly small upon passivation. Cr atom doping generates acceptor levels in the bare cluster along with deep levels whereas in the passivated cases it generates deep levels and donor levels in CR substitution. Mn atom doping leads to acceptor levels in the system except for the passivated CR site doped cluster. States near the HOMO have dominant contribution from Te states which disappears on passivation for CR substitution. Unlike CR, in OR, upon passivation the acceptor levels are seen along with deep levels which are mainly due to Cd and Te. In case of Fe, in unpassivated cluster, the acceptor levels are mainly due to Te with small contributions from Cd and Fe. It also forms deep levels. On passivation, acceptor levels appear near the HOMO with enhanced Fe contribution.

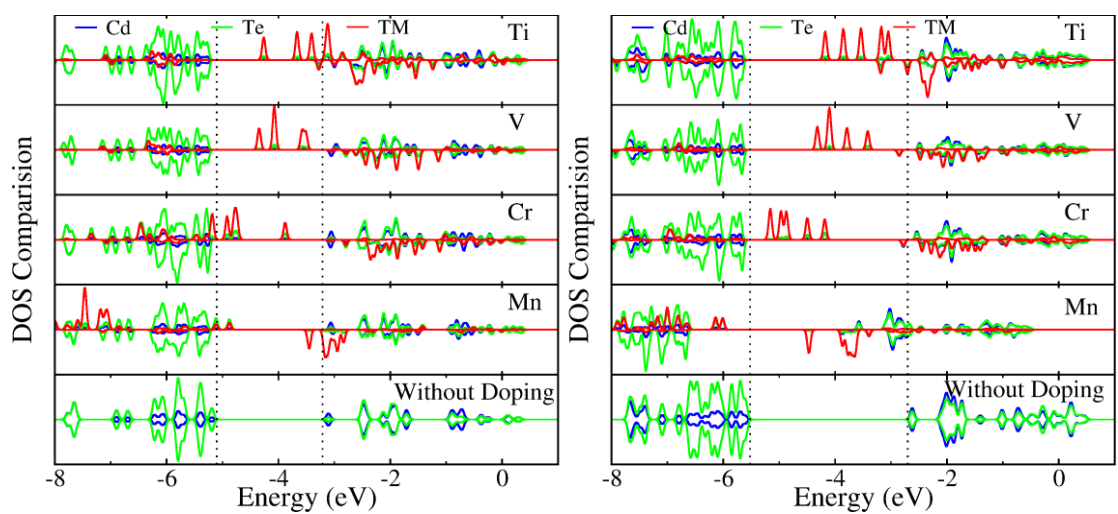

Figure 6. PDOS comparison of 3d TM (Ti, V, Cr, Mn) atom doped at CR site in bare and FHPA passivated $Cd_9Te_9$ cluster by mapping on energy scale of undoped cluster's PDOS.

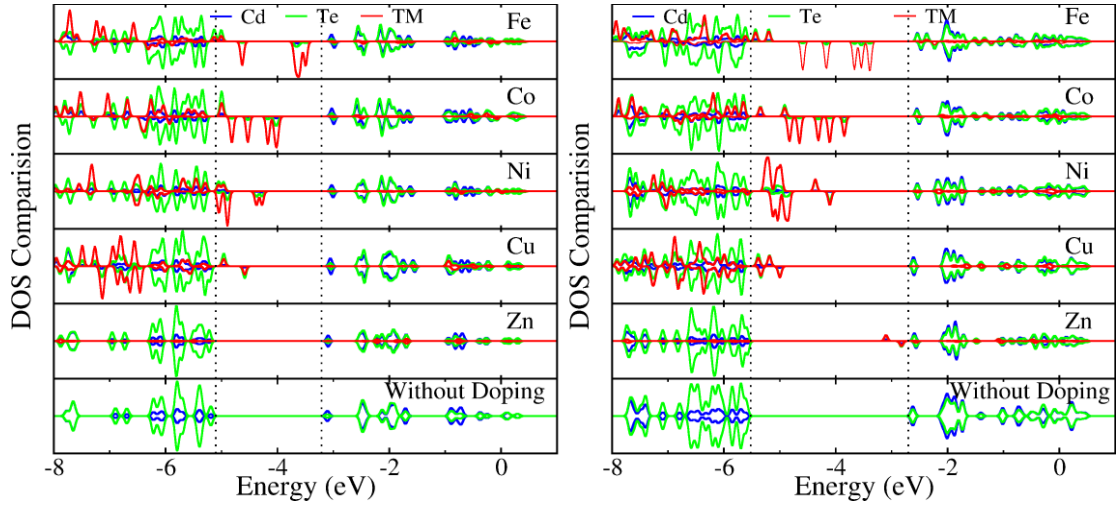

Figure 7. PDOS comparison of 3d TM (Fe, Co, Ni, Cu, Zn) atom doped at CR site in bare and FHPA passivated $Cd_9Te_9$ cluster by mapping on energy scale of undoped cluster's PDOS.

Co atom doping generates acceptor (mainly due to Te) and deep levels in the cluster. Similarly, Ni contributes acceptor and deep levels except the OR passivated cluster case. Polarity of deep level states does not change upon passivation. Cu atom doping leads to formation of acceptor levels and those are retained even after passivation. These levels have major contribution due to Te. Zn atom doping does not lead to any states in the energy gap of host cluster in bare cluster cases. However, donor levels appear on passivation of cluster in CR site doping case.

Overall, Ti, Mn and Fe doping results into electron type doping in bare clusters. Zn and V atom doping also results in electron doping in passivated cluster with CR site doping. In other cases, except for V (which forms deep levels), all are hole doped systems. States due to the Te-p orbitals in the gap region are shifted out of the band gap region upon passivation as a result, p-d hybridization is weakened.

We have also studied the effect of doping few 4d transition metal atoms (Ru, Rh, Pd) for a Cd in OR and CR and the results are presented below. Figure 8 (S7 in SI) shows the partial density of states for 4d-TM (Ru, Rh, Pd) doped clusters in OR (CR) for bare and FHPA passivated $Cd_9Te_9$ cluster.

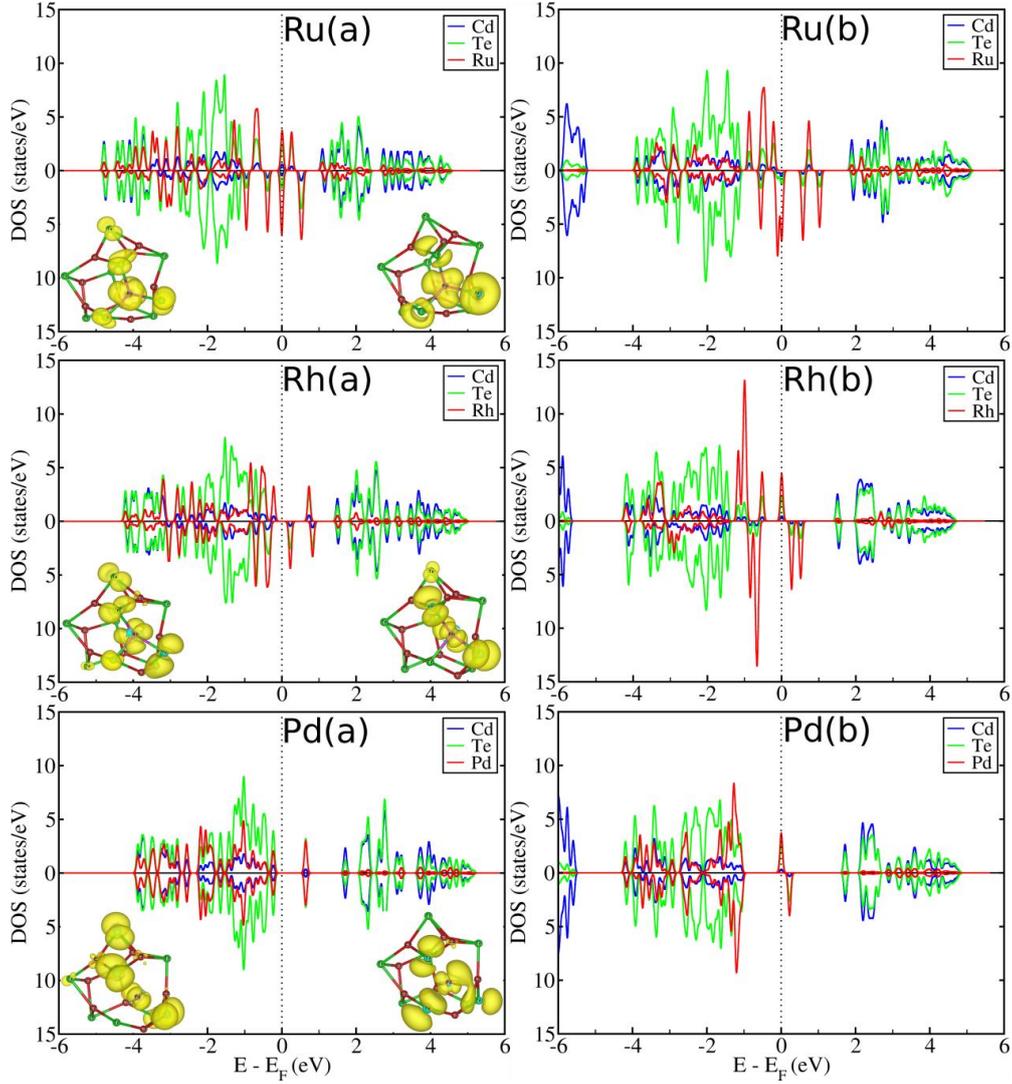

Figure 8. Site projected density of states for TM atom doped (a) bare and (b) FHPA passivated $Cd_9Te_9$ clusters for different 4d TM (Ru, Rh, Pd) atom substitution for a Cd in OR. Fermi level ($E_F$), denoted by the vertical dotted line, has been shifted to 0 eV. For semiconducting systems, the electronic charge densities corresponding to HOMO (left of $E_F$) and LUMO (right of $E_F$) are given as inset in corresponding figures.

Ru atom substituted for Cd atom in OR of bare $Cd_9Te_9$ cluster has finite states at $E_F$, thus it is metallic in nature, however, the CR doped cluster has a small energy gap of 0.2 eV. On passivation, both the systems show half-metallic nature due to minority charge carriers. The states in HOMO appear as a result of hybridization of Cd-s, Te-p and Ru-d states for OR case. Upon passivation, the states due to the minority carriers are retained while those due to majority carriers vanish, which results into half-metallic nature of system. The CR doped cluster has p-d hybridized LUMO.

Unlike Ru, Rh atom doping does not change the semiconducting nature of host system but there is reduction in band gap. On passivation, OR doped cluster acquires half-metallic nature due to majority charge carriers. In case of passivated CR doped cluster, there are conducting states due to minority charge carriers at $E_F$. Therefore, which carrier will form the conducting channel is related to the physical environment around the dopant atom. HOMO in both the cases has predominant contribution due to Rh-d and Te-p, whereas the LUMO has s-d hybridized states along with Te-p.

Pd atom doped bare cluster is seen to be semiconducting with equal band gaps for both the majority and minority charge carriers. On passivation when it becomes a half-metal, the CR doped cluster has higher semiconducting gap due to minority charge carriers. PDOS of the Pd doped bare $Cd_9Te_9$ cluster, including the states due to Pd, is symmetric about the zero-axis and results into a nonmagnetic system. On passivation, there is uneven shift in the states from above the Fermi level towards the Fermi level which leads to half-metallic nature of system and it also develops a small magnetic moment of 0.2 $\mu_B$. Pd has filled 4d states but the 5s states are empty. This electronic configuration difference between Pd and Zn atoms is thus reflected in the corresponding PDOS.

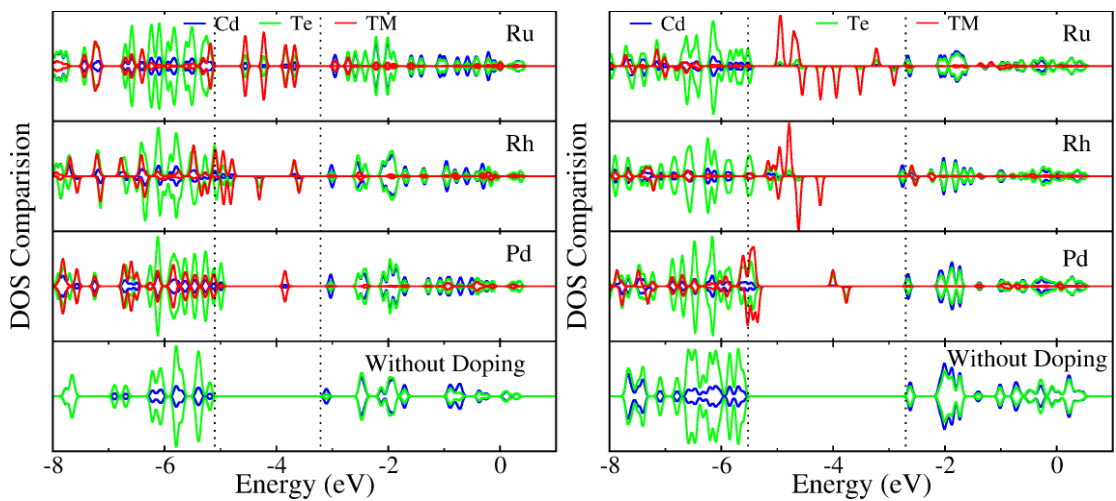

Figure 9. PDOS comparison of 4d TM (Ru, Rh, Pd) atom doped for a Cd at CR site in the bare and FHPA passivated $Cd_9Te_9$ cluster by mapping on energy scale of undoped cluster's PDOS.

The PDOS of all the 4d-TM systems with and without passivation are plotted without shifting the zero of energy axis to Fermi level for comparison in figures 9 and S6 in SI. For all three cases there is effective doping of acceptor levels (except Ru in bare cluster case of doping at CR site) in the system along with deep levels. In the Ru doping case, there is uniform distribution of levels in energy gap region. In all the three cases, the states due to TM increase in the gap region upon passivation. For Ru and Rh, in passivated CR doping case, it has the acceptor, deep and donor levels formed. The acceptor levels due to Rh atom doping have major Te contribution which decreases on passivation whereas, in passivated-CR case, donor levels, mainly due to Cd and Te, are present.

The reduction in the spatial dimension of a semiconductor material is accompanied by an increment in the band gap of the material (blue shift) due to quantum size effect. Different theories viz. the effective mass approximation, empirical psuedopotential calculations, tight binding models etc. are used to understand the underlying physics of this phenomenon. The contribution to the enhancement in the band gap is not symmetric across the Fermi level as it has been shown that the shift in the edge of HOMO is much larger than that in the LUMO and is inversely proportional to the effective mass of electron in that band [**34**]. For small semiconducting clusters, the electronic structure and its geometry are strongly correlated [**35**].

The quantum effects become dominant over the surface effects when the system size becomes smaller than the excitonic radius.

In our studies, we have the maximum diameters of undoped, bare and passivated $Cd_9Te_9$ clusters as 7.86 Å and 8.32 Å, respectively. Passivation of $Cd_9Te_9$ with FHPA leads to significant changes in the geometry affecting the stability and the electronic structure viz. increase in the binding energy and enhancement of energy gap. The average bond lengths of the Cd-Te bonds increase as result of the reduced Te-Te Coulomb repulsion, upon passivation. 5p states of Te happen to be the outermost i.e., the surface states of the cluster and since they are partially filled, they predominantly contribute to HOMO and LUMO. When the cage is mono-doped by a set of 3d and 4d transition metal atoms which can introduce finite magnetic moment(s) in the system, for a Cd in OR and CR, the electronic structure is changed to a large extent, accompanied by small changes in geometry of the cluster. The geometries of bare and passivated clusters for the cases of OR and CR are shown in Figure S1 of SI. In the CR doped passivated clusters, TM atom(s) with partially filled d states attracts the FHPAs bonded to nearest Te atoms. This leads to the distortion in the geometry more than the other cases. Table ST1 gives the average TM-Te bond length for all the clusters studied in the preset work. It shows that Te atoms move closer to the TM atom as compared to Cd atom(s) in the undoped case. When a TM atom replaces a Cd atom in CR or OR, relaxed structure shows the movement of TM towards the interior of cage. This movement towards the interior results in the reduction in average TM-Te bond-length in doped cluster from Cd-Te in undoped cluster as can be seen in Table ST1. The Te atoms considered for measurement of bond-length are the nearest neighbors of TM (NNTM). The average TM-Te bond-lengths are near to each other for doping of Ti, V, Cr, Mn cases. Beyond which it falls with atomic number till Ni. For Cu and Zn, it remains in the previous range. The lowest average TM-Te bond-length happens to be for Ni doped cluster. In 4d TM doping cases, the average TM-Te bond-lengths is observed to increase with atomic number (exception Pd doping for a Cd in CR of passivated cluster).The reduction in TM-Te bond lengths for passivated case is more as compared to the unpassivated cases. The average variation in the Cd-Te bond-length in the ring where the TM is doped is seen to have reduced as compared to that in the undoped cluster. This reduction is compensated to a certain extent by the accompanying increment in the Cd-Te bond length in the other rings.

The magnetic moment does not change significantly with the position of TM doping for Cd, except for Ni and 4d elements. The magnetic moments on the TM atom(s) in the cluster are given in table 1 and are also plotted in figure 10.

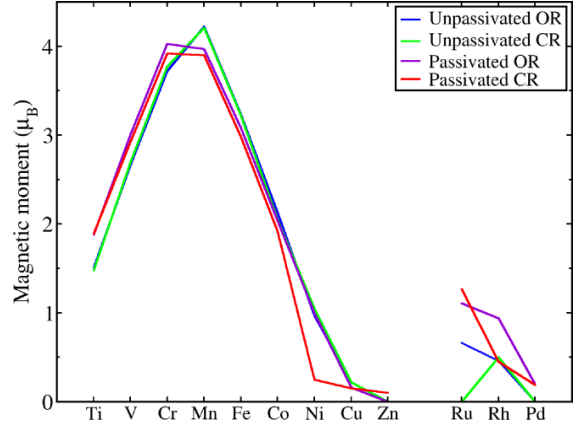

Figure 10. Magnetic moments on TM atom, doped for a Cd at OR and CR site of bare and passivated $Cd_9Te_9$ clusters.

| TM | Magnetic Moments($\mu_B$) | | | |
| --- | --- | --- | --- | --- |
| | Without passivation | | On passivation | |
| | OR | CR | OR | CR |
| Ti | 1.51 | 1.48 | 1.88 | 1.89 |
| V | 2.66 | 2.69 | 3.00 | 2.92 |
| Cr | 3.72 | 3.77 | 4.03 | 3.92 |
| Mn | 4.22 | 4.21 | 3.97 | 3.90 |
| Fe | 3.23 | 3.22 | 3.08 | 2.98 |
| Co | 2.13 | 2.05 | 2.04 | 1.92 |
| Ni | 0.97 | 1.05 | 1.00 | 0.25 |
| Cu | 0.22 | 0.22 | 0.16 | 0.15 |
| Zn | 0.00 | 0.00 | 0.00 | 0.10 |
| Ru | 0.66 | 0.00 | 1.11 | 1.27 |
| Rh | 0.46 | 0.50 | 0.94 | 0.45 |
| Pd | 0.00 | 0.00 | 0.20 | 0.19 |

Table 1: Magnetic moments on TM atom, doped for a Cd at OR and CR site of bare and passivated $Cd_9Te_9$ clusters.

In the unpassivated clusters, TM doping results in states in the energy gap region of host cluster with p-d hybridization. Also, the polarized states result in finite magnetization. The magnetic moment is maximum (4.2 $\mu_B$) on the Mn atom and zero for the case of Zn, Pd and the CR doping case of Ru. Presence of dangling bonds alters the p-d hybridization and as a result, these systems retain the semiconducting nature. The minimum energy gap of 0.1 eV is observed for Co and Cu doping cases. But, on passivation of the cluster using FHPA, the semiconducting system exhibits half-metallic nature. The half-metallic nature can be due to the majority or minority carriers depending on the dopant atom and physical environment around it. For example, the majority carriers are conducting for the Ti, V Cr doped cases, whereas the Mn, Fe, Co and Cu doping leads to minority conduction. In addition to this, couple of elements have a dependence on the local physical environment. In the CR doped case, the Ni atom has the majority carriers conducting whereas in OR doping, it is the minority carriers. For the 4d elemental doping, the cluster exhibits majority conduction in Rh (OR) and Pd whereas for Ru and Rh (CR), it is the minority charge carriers. The charge density corresponding to the HOMO and LUMO in these case, are hybridized. The $dz^2$ orbital due to TM happens to be predominant in the LUMO for Ti, V, Cr and Mn, while it is in HOMO for Fe in OR doping. This behaviour is attributed to the extent of the hybridization in LUMO for Ti, V, Cr and Mn. But, for CR doping case, LUMO of Ti and HOMO of Fe are mainly due to TM-$dz^2$. The doped TM atom is surrounded by the Te atoms which then results in the hybridization of 5p orbitals of Te with the 3d and 4d of the TM leading to the p-d hybridization seen in these systems.

Substitutional TM doping in $Cd_9Te_9$ results in introduction of different levels in the energy gap region. Therefore, each system can be viewed as doped with holes (acceptor levels), electrons (donor levels) and/or the deep levels in the HOMO-LUMO gap. The TM d states are also seen deep in the valence region or up in the conduction region. (Table ST3 in SI summarizes the individual doping behavior). For all the 4d elements studied here, it is akin to the doping with holes, with the exception (electron-hole doping) for Ru and Rh in passivated case of CR. In the 3d elemental doping, Ti, Mn and Fe doping results into the electron type doping in bare cluster whereas Ti, V, Zn and Ru in the passivated CR site doped cases. Most of elements dope acceptor levels except the V atom (which forms only the deeps levels).

**Conclusions**

A stable cage-like geometry is obtained for $Cd_9Te_9$ cluster after geometry optimization is performed. The electronic structure calculations are then carried out and the structural changes have been shown to accompany the electronic structure changes and are directly related to the repulsive Coulomb interactions among the Te atoms. Further, these bare and FHPA passivated clusters are doped with some 3d and 4d TMs for understanding their magnetic and electronic behaviour. Two non-equivalent sites are used for the substitutional mono-doping derived from the symmetry of the clusters. All the unpassivated mono-doped clusters are semiconducting in nature, while mono-doping in FHPA passivated clusters induces the system to become half-metallic in nature. For the 3d TM doping cases, as expected, the magnetic moment increases from Ti to Mn as the number of electrons in the d orbital increases whereas it decreases from Mn to Zn since the number of d electrons in the down spin state increases with the up spin states completely filled. This observation agrees with the Slater-Pauling magnetization curve. All the dopants show an increase in the magnetic moment upon passivation for the case of 4d TM doping. The doping of TM for Cd induces energy levels in the HOMO-LUMO gap but the position and distribution of these states in the gap is different. Ti, Mn and Fe dope electrons independent of physical environment or passivation whereas Ru, Rh and Pd dope electrons when doped in CR upon passivation. Most of the TM atoms introduce hole-like or acceptor states in the gap. Cr and Mn doping results in highest magnetic moment for the system with the up and down spin states separated fairly well in energy. These systems are therefore suitable candidates for spintronic applications. Self-assembly of such stable structures to form quasi one-dimensional or quasi two-dimensional structures can provide a plausible route for innovative device applications.

**Acknowledgement**: KTC acknowledges DAE/IGCAR for the research fellowship and AK acknowledges DST Nanomission Council for financial support through a major research project to set up a high performance computing facility (DST/NM/NS-15/2011(G)).

# Supplementary Information for

# Half-Metallicity in Smallest Cage-like Cluster of CdTe with Doping of Transition Metal Atoms

Kashinath Tukaram Chavan[a], Sharat Chandra*[,a] and Anjali Kshirsagar[b]

[a]*Materials Physics Division, Materials Science Group, Indira Gandhi Centre for Atomic Research, HBNI, Kalpakkam 603 102, INDIA*

[b]*Department of Physics, Savitribai Phule Pune University, Pune 411 007, INDIA*

*\*Corresponding author email: sharat@igcar.gov.in*

Table ST1. Average distance of TM from the nearest three Te atoms in TM doped for Cd in CR and OR site of bare and passivated cluster $Cd_9Te_9$ cluster.

| TM | Average TM-Te distance (Å) | | | |
|---|---|---|---|---|
| | OR | | CR | |
| | Bare | Passivated | Bare | Passivated |
| Cd | 2.82 | 2.86 | 2.82 | 2.86 |
| Ti | 2.72 | 2.76 | 2.71 | 2.74 |
| V | 2.66 | 2.71 | 2.69 | 2.68 |
| Cr | 2.67 | 2.71 | 2.70 | 2.70 |
| Mn | 2.68 | 2.67 | 2.69 | 2.67 |
| Fe | 2.59 | 2.60 | 2.60 | 2.58 |
| Co | 2.54 | 2.54 | 2.53 | 2.54 |
| Ni | 2.50 | 2.51 | 2.52 | 2.48 |
| Cu | 2.54 | 2.54 | 2.58 | 2.58 |
| Zn | 2.63 | 2.65 | 2.63 | 2.70 |
| Ru | 2.52 | 2.57 | 2.53 | 2.60 |
| Rh | 2.55 | 2.58 | 2.61 | 2.65 |
| Pd | 2.59 | 2.60 | 2.65 | 2.62 |

Table ST2. The HOMO-LUMO gap for $Cd_9Te_9$ cluster on mono-doping by TM atoms for Cd in OR and CR, with and without passivation. Up (down) spin electrons are referred as majority (minority) charge carriers.

| TM | Substitution for Cd in OR | | | | Substitution for Cd in CR | | | |
|---|---|---|---|---|---|---|---|---|
| | HOMO-LUMO gap (in eV) | | | | HOMO-LUMO gap (in eV) | | | |
| | Without passivation | | With passivation | | Without passivation | | With passivation | |
| | Majority Spin | Minority Spin | Majority Spin | Minority Spin | Majority Spin | Minority Spin | Majority Spin | Minority Spin |
| Ti | 0.25 | 1.7 | - | 2.2 | 0.1 | 1.8 | - | 2.3 |
| V | 0.2 | 2.1 | - | 2.6 | 0.2 | 2.2 | - | 2.8 |
| Cr | 0.4 | 2.2 | - | 2.8 | 0.7 | 2.0 | - | 2.8 |
| Mn | 1.2 | 1.2 | 2.6 | - | 1.7 | 1.7 | 2.3 | - |
| Fe | 1.8 | 0.6 | 2.7 | - | 1.8 | 0.7 | 2.6 | - |
| Co | 1.7 | 0.1 | 2.4 | - | 1.8 | 0.1 | 2.3 | - |
| Ni | 1.2 | 0.2 | 2.3 | - | 1.8 | 0.3 | - | 0.55 |
| Cu | 1.7 | 0.1 | 2.3 | - | 1.7 | 0.45 | 2.4 | - |
| Zn | 1.8 | 1.8 | - | - | 1.9 | 1.9 | - | 2.6 |
| Ru | - | - | 0.7 | - | 0.2 | 0.2 | 1.25 | - |
| Rh | 0.7 | 0.3 | - | 0.7 | 1.0 | 0.3 | 1.8 | - |
| Pd | 0.7 | 0.7 | - | 1.1 | 1.0 | 1.0 | - | 1.5 |

Table ST1 lists the HOMO-LUMO gaps for majority and minority charge carriers. The actual gap can be different if it is measured across the spin degree of freedom. Ru doping for Cd in OR leads to a transition from semiconductor to metal in unpassivated cluster, whereas for all other TM atom(s) doping, semiconducting nature is retained. Zn atom

doped system retains its metallic nature in all the situations except substitution in CR on passivation. All other systems undergo transition from semiconductor to half-metal on passivation, however the conducting states are due to different types of carriers, viz., majority charge carriers have non-zero states at Fermi level for Ti, V, Cr, Rh, and Pd (Ti, V, Cr, Ni, Zn and Pd) while for rest of the TM atoms i.e., Mn, Fe, Co, Ni, Cu and Ru (Mn, Fe, Co, Cu, Ru and Rh) minority charge carriers have non-zero states at Fermi level in the OR (CR) site doping. For Ni doping, switching for conducting states from minority carriers to majority carriers is observed when the doping site is switched from OR to CR.

Figure S1 depicts the geometries of TM doped bare and FHPA passivated clusters. If the cluster is viewed in a perspective such that it appears to be comprised of three (parallel when passivated) rings, then the outer two rings are identical. As a Cd atom in the outer ring of $Cd_9Te_9$ cluster is replaced by TM (as shown in figure S1(a)), it has influence on the positions of atoms in its vicinity, i.e., nearest and next nearest atoms. The distortion in geometry is more is the passivated case. When the TM atom is doped in CR, as shown in figure S1(b), it has identical physical environment as far as the Te atoms in outer ~~central~~ ring are concerned. But the Te atoms in the central ~~outer~~ rings are not placed symmetrically. Hence presence of TM atom leads to distortion in central ring. As the TM moves below the plane of three Te atoms, towards the cage interior, the chair gets distorted.

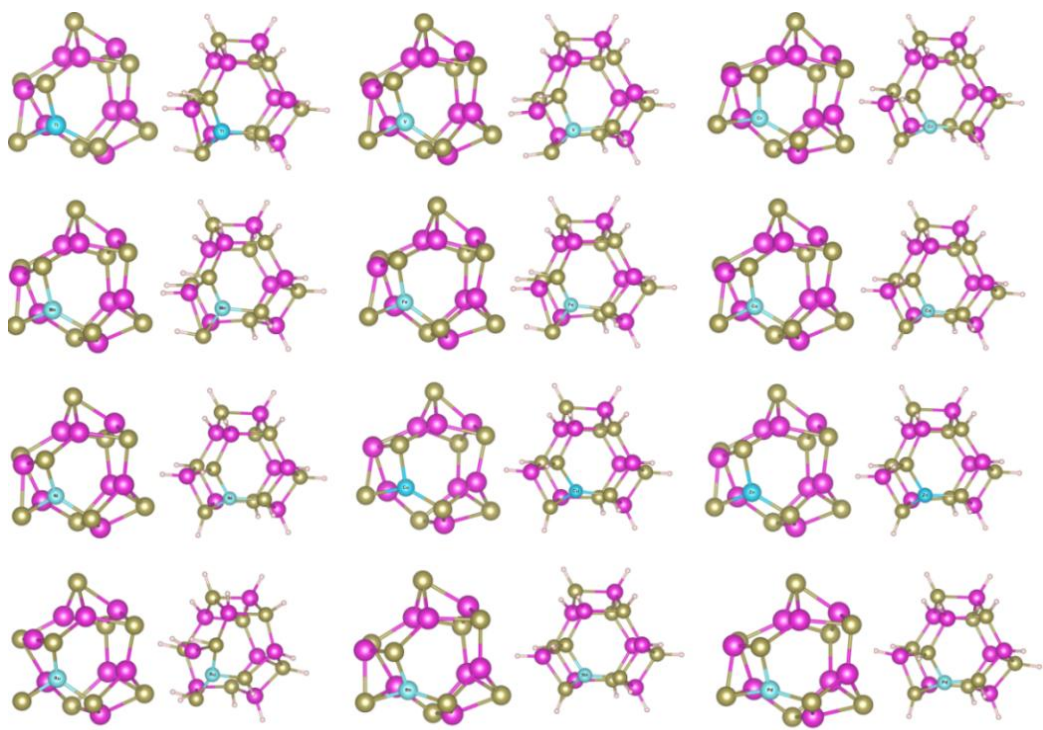

(a)

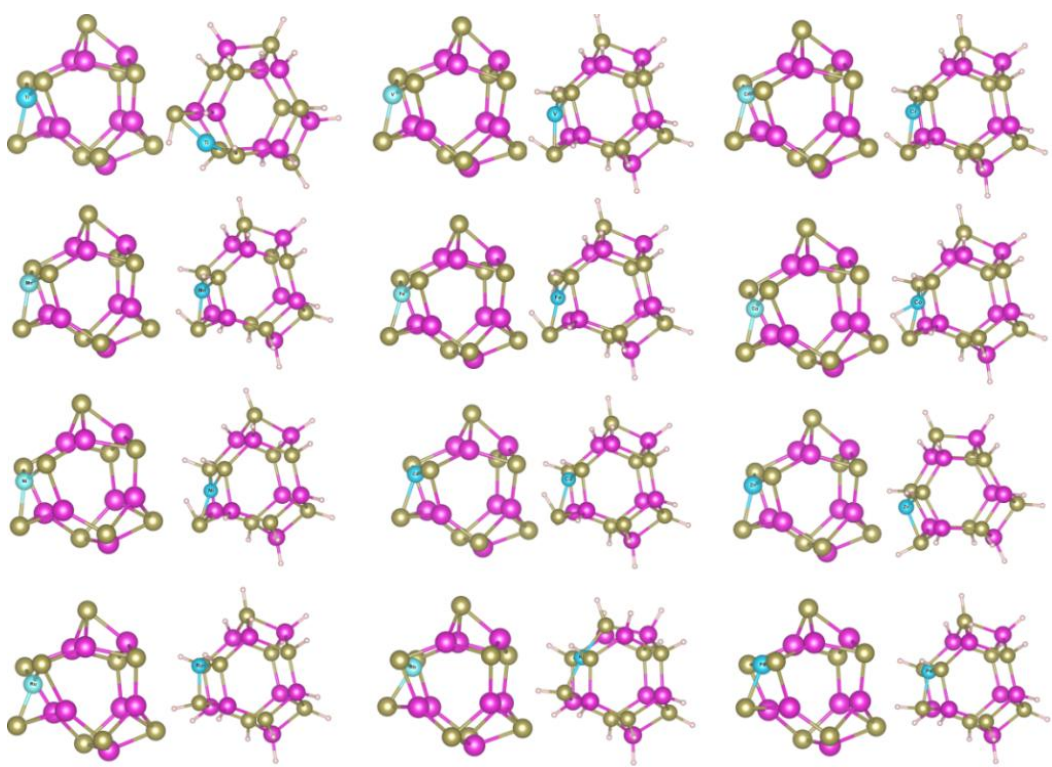

(b)

Figure S1. Geometries of TM atom(s) doped for Cd at (a) OR site and (b) at CR site in bare and FHPA passivated $Cd_9Te_9$ cluster. The magenta and brown spheres are Cd and Te atoms respectively and small gray atoms are FHPA. The TM (Ti, V, Cr, Mn, Fe, Co, Ni, Cu, Zn, Ru, Rh and Pd) atoms are blue spheres and are shown from left to right and top to bottom rows in that order respectively.

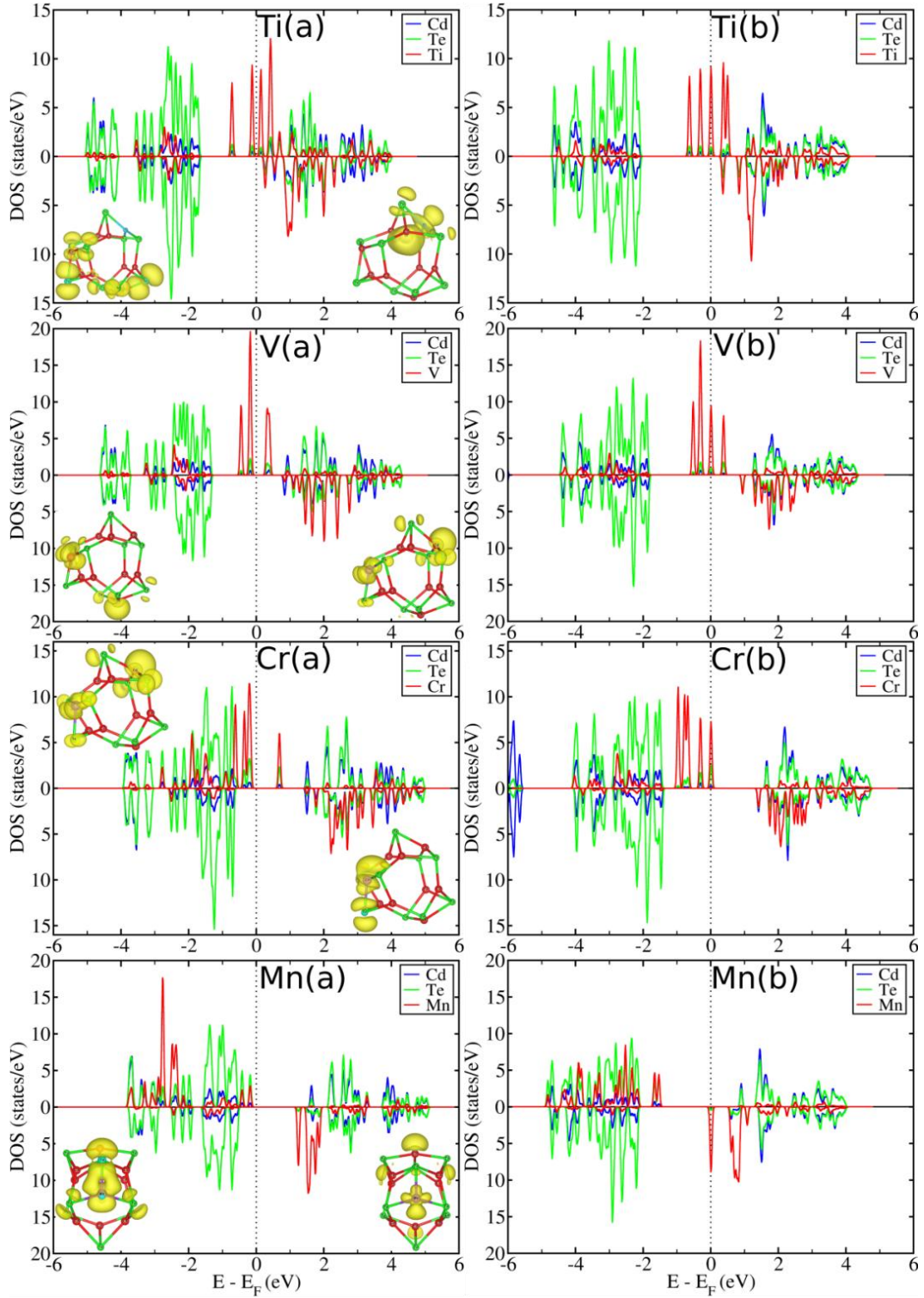

Figure S2. Site projected density of states for TM atom doped (a) bare and (b) FHPA passivated $Cd_9Te_9$ clusters for different 3d TM (Ti, V, Cr, Mn) atom substitution for Cd in the CR. Fermi level ($E_F$), denoted by the vertical dotted line, has been shifted to 0 eV. The bare doped clusters are semiconducting in nature, their electronic charge densities corresponding to HOMO (left of $E_F$) and LUMO (right of $E_F$) are shown as inset in corresponding figures. These charge densities are plotted at a common isovalue of 0.002 at which the orbital characters are visible. This isovalue ranges from 10% to 20% of maximum value of partial charge density in individual case.

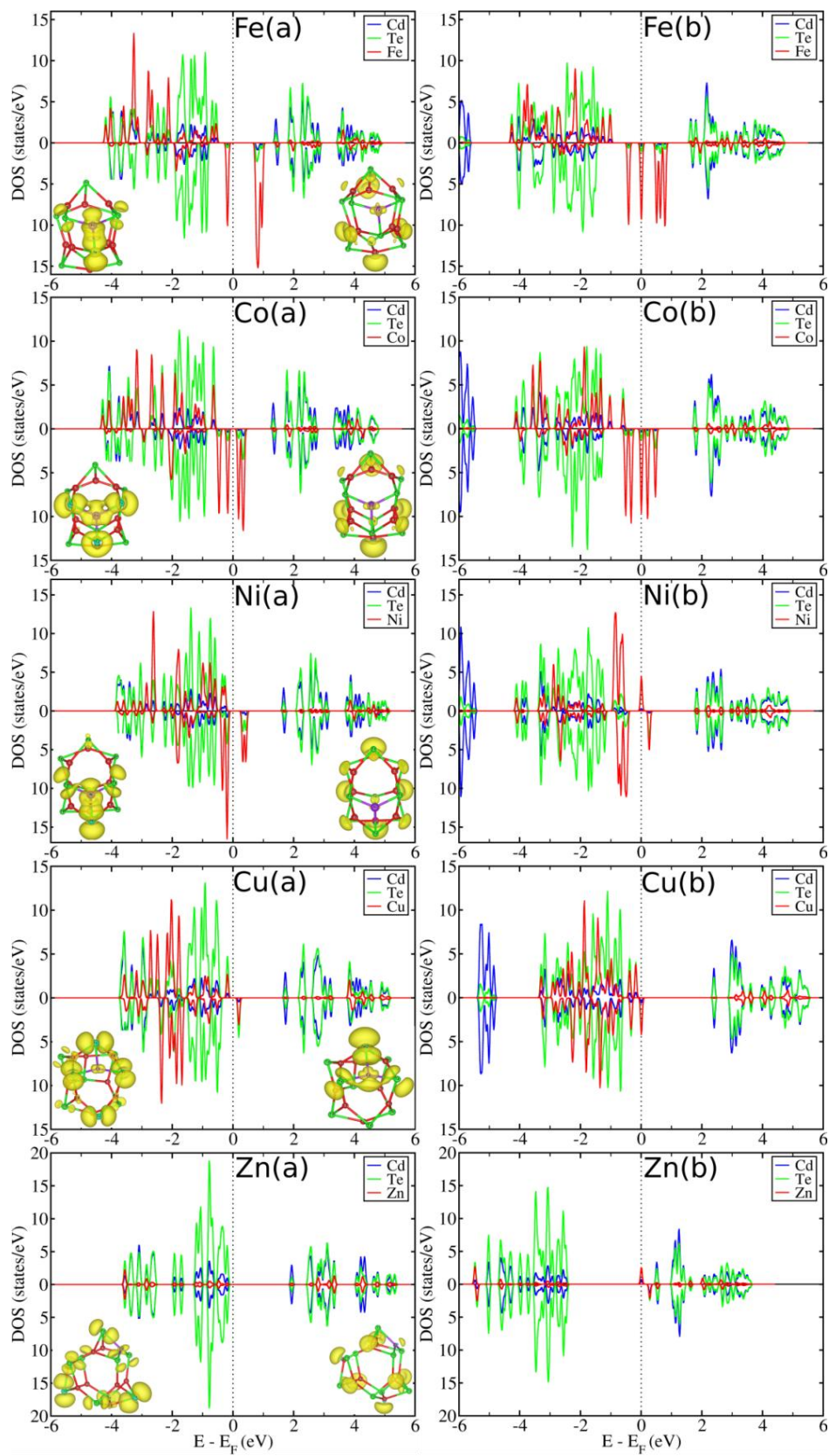

Figure S3. Site projected density of states for TM atom doped (a) bare and (b) FHPA passivated $Cd_9Te_9$ clusters for different 3d TM (Fe, Co, Ni, Cu, Zn) atom substitution for Cd in the CR. $E_F$, denoted by the vertical dotted line, has been shifted to 0 eV. The bare doped clusters are semiconducting in nature, their electronic charge densities corresponding to HOMO (left of $E_F$) and LUMO (right of $E_F$) are shown as inset in corresponding figures.

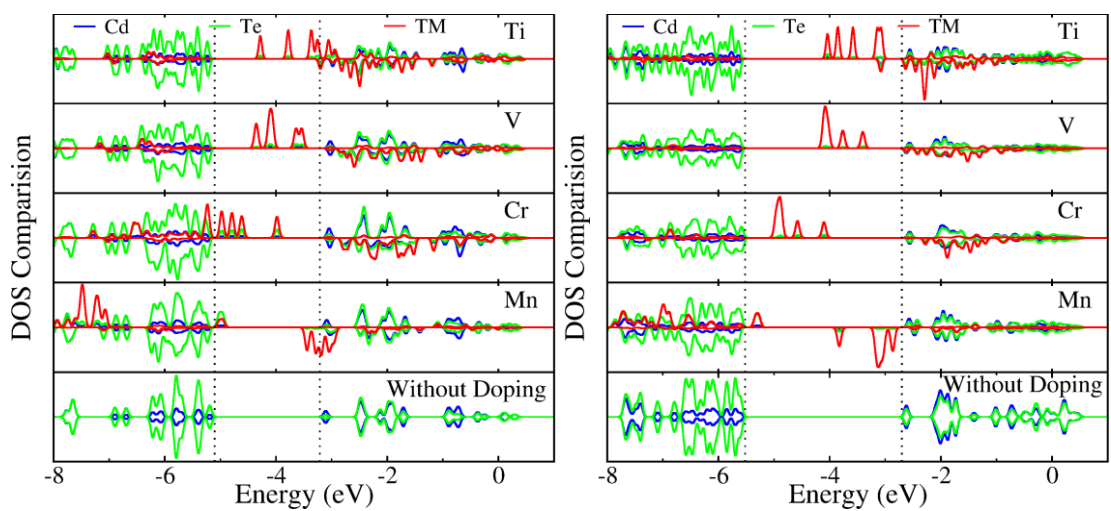

Figure S4. PDOS comparison of 3d TM (Ti, V, Cr, Mn) atom doped at OR site in bare and FHPA passivated $Cd_9Te_9$ cluster by mapping on energy scale of undoped cluster's PDOS.

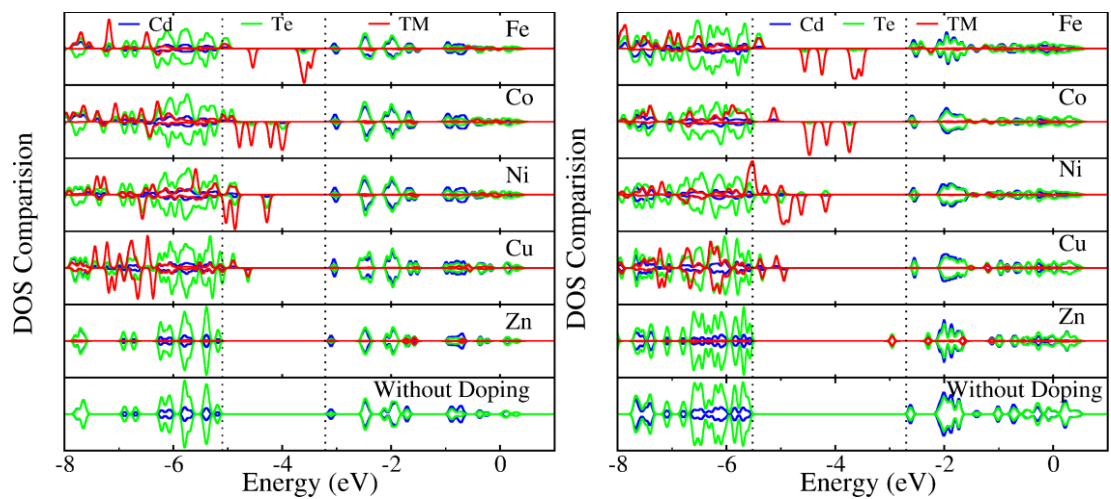

Figure S5. PDOS comparison of 3d TM (Fe, Co, Ni, Cu, Zn) atom doped at OR site in bare and FHPA passivated $Cd_9Te_9$ cluster by mapping on energy scale of undoped cluster's PDOS.

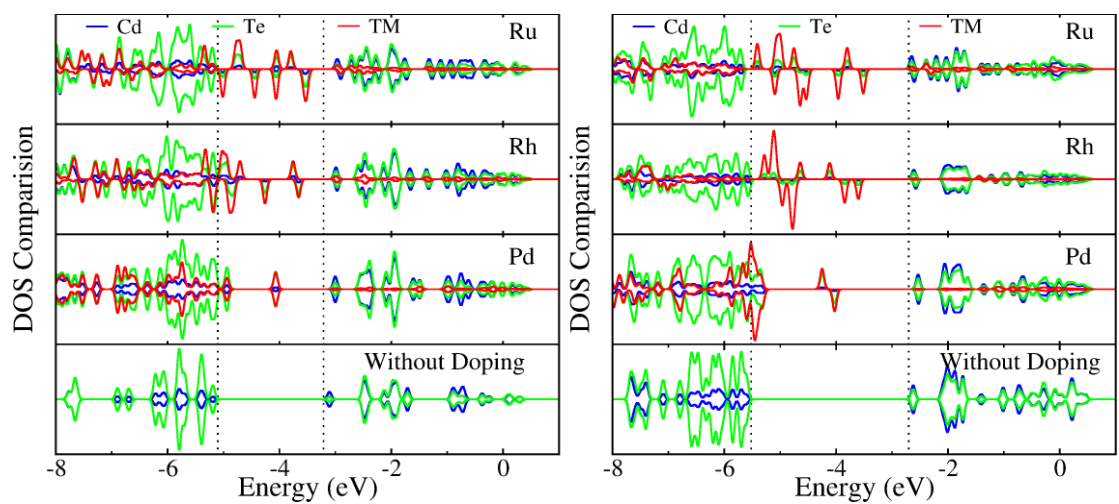

Figure S6. PDOS comparison of 4d TM (Ru, Rh, Pd) atom doped for a Cd at OR site in bare and FHPA passivated $Cd_9Te_9$ cluster by mapping on energy scale of undoped cluster's PDOS.

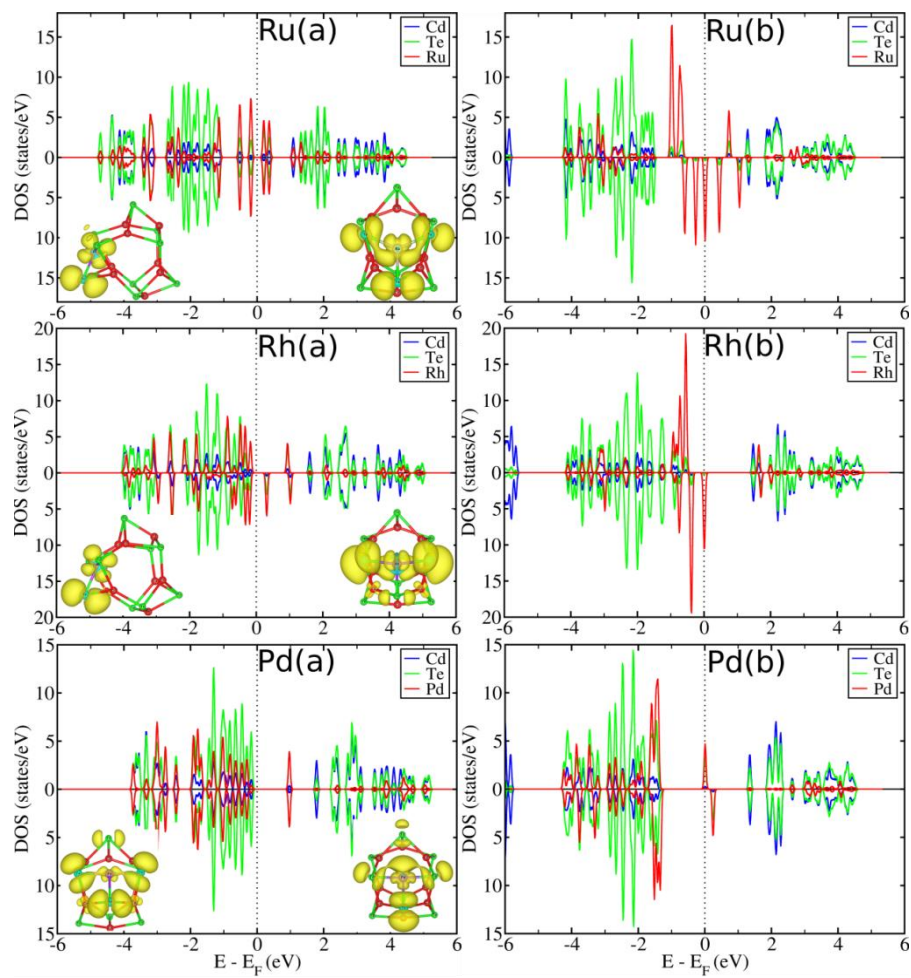

Figure S7. Site projected density of states for TM atom doped (a) bare and (b) FHPA passivated $Cd_9Te_9$ clusters for different 4d TM (Ru, Rh, Pd) substitutions for a Cd in CR. $E_F$, denoted by the vertical dotted line, has been shifted to 0 eV. For clusters with semiconducting nature, the electronic charge densities corresponding to HOMO (left of $E_F$) and LUMO (right of $E_F$) are given as inset.

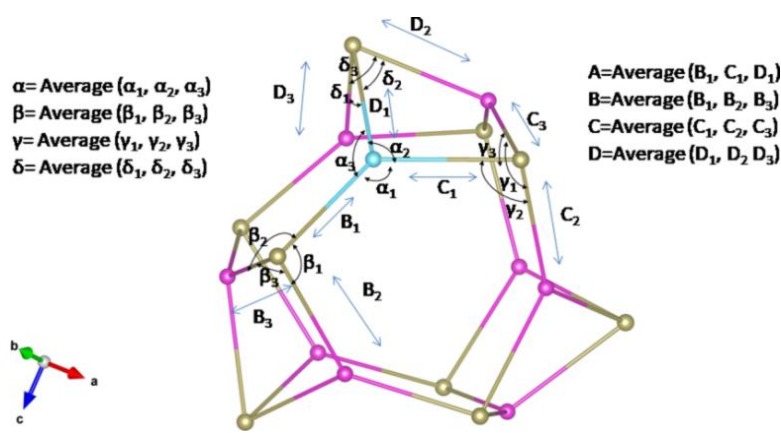

Figure S8. A model for bond-lengths and bond-angles. A TM (blue) atom is doped for Cd (magenta) atom in outer ring of unpassivated $Cd_9Te_9$ cluster. Brown spheres are Te atoms. Same color scheme is followed in all the remaining figures. A is average TM-Te (nearest) bond-length and α is average Te-TM-Te angle. Similarly the average bond-lengths and bond-angles with respect to Te atoms (3 Te nearest to TM) are B, C, D and β, γ, δ etc. These bond angle and bond-length definitions may change according to the TM doping position. i.e. OR and CR as shown in figures below.

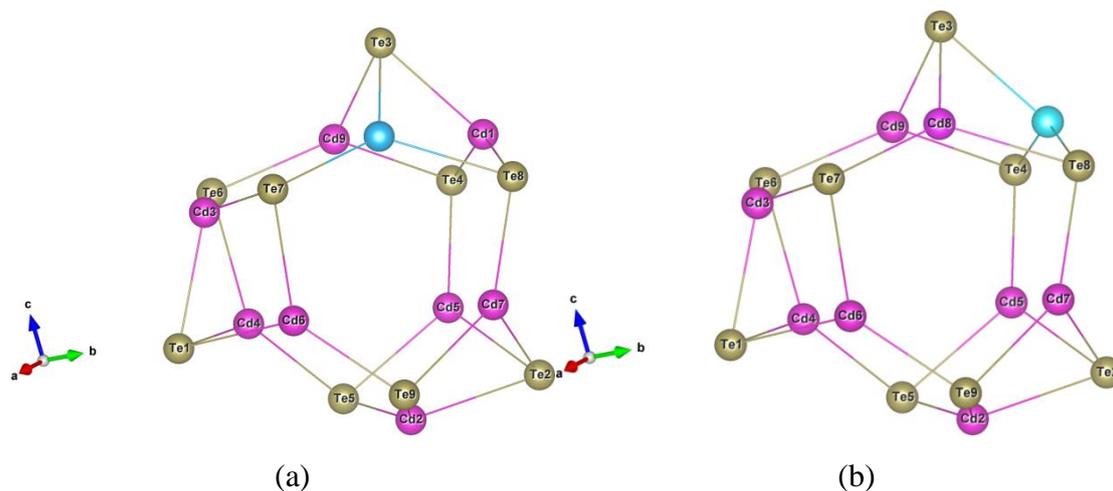

(a)            (b)

Figure S9. TM atom substituted for a Cd atom (a) in OR and (b) in CR. The average bond-lengths between TM and nearest Te (Te3, Te7, Te8 in (a) and Te3, Te4, Te8) in (b)) are denoted as A and bond-angles between those Te are denoted as α. Similarly, for the nearest three Te atoms in the respective figures the average bond-lengths and bond-angles are represented as B, C, D and β, γ, δ.

For the case of TM atom doping at OR of unpassivated cluster (as shown in figure S9(a)), the average Te-TM-Te bond-angles are close to that of Te-Cd-Te in undoped case. Also, the bond-angles with respect to Te atom(s), angles β and γ, are close to those in undoped cases, except for the case of Ru atom doping. Average bond angles δ are also not affected except for the 4d TM atoms. For Ru doping, there are significant changes in the geometry. The reduction in average bond-lengths as compared to the undoped cases is due to a reduction in the TM-Te bond-length(s), otherwise Cd-Te bond-lengths are least affected. For the 4d TM atom(s) doping, a Cd atom at equivalent position (in parallel) on other outer ring moves towards the TM atom(s), and hence reduction in TM-Te3-Cd9 angle is reflected in the value(s) of angle δ. Thus, the increment in this angle as a function of atomic number suggests a reduction in TM-Cd attraction.

Figure S9(b) is a model for the bond-lengths and bond-angles for systems with TM atom(s) substituted at the position of a Cd in CR in unpassivated bare clusters. As there are identical rings on both sides of TM atom, the observed bond-lengths and bond-angles in symmetric directions are equal. The Te atom in central ring i.e., Te3 moves towards the TM, more than the other two (Te4 and Te8), hence its contribution in reduction of TM-Te average bond-lengths is more. Although there are changes in Te-TM-Te bond-angles, on an average, they are close to the Te-Cd-Te (at identical position(s)) angles of the undoped cluster. More changes are observed for 4d TM atom doping cases, especially for Ru and Rh. TM atom moves inward with Te following it whereas the Cd atoms (Cd8, Cd9) bonded to Te (Te3) on central ring move away from the TM atom. Hence the outer hexagonal rings in 4d atom doped systems are more distorted.

The definitions of the various bond-lengths and the bond-angles for TM atom doped passivated $Cd_9Te_9$ cluster are shown in figure S10. The average TM-Te bond lengths are nearly equal for V and Cr atom doping in OR of passivated cluster, but then decrease from Mn with increase in atomic number for 3d TM atom doped cases. The average bond lengths are least affected for the case of Ti doping. In contrast to most of the 3d TM atom doped systems, average bond-lengths increase for 4d TM atom doped systems. The Te-TM-Te bond angle close to 120° reveals that the TM atom is in the plane containing the next near neighbor Te atoms. The bond-lengths B, C and D also show a trend similar to the TM-Te bond-lengths A for 3d TM doped cases. The geometry of the cluster is more distorted for Ru atom doping as reflected in the respective bond-lengths and bond-angles.

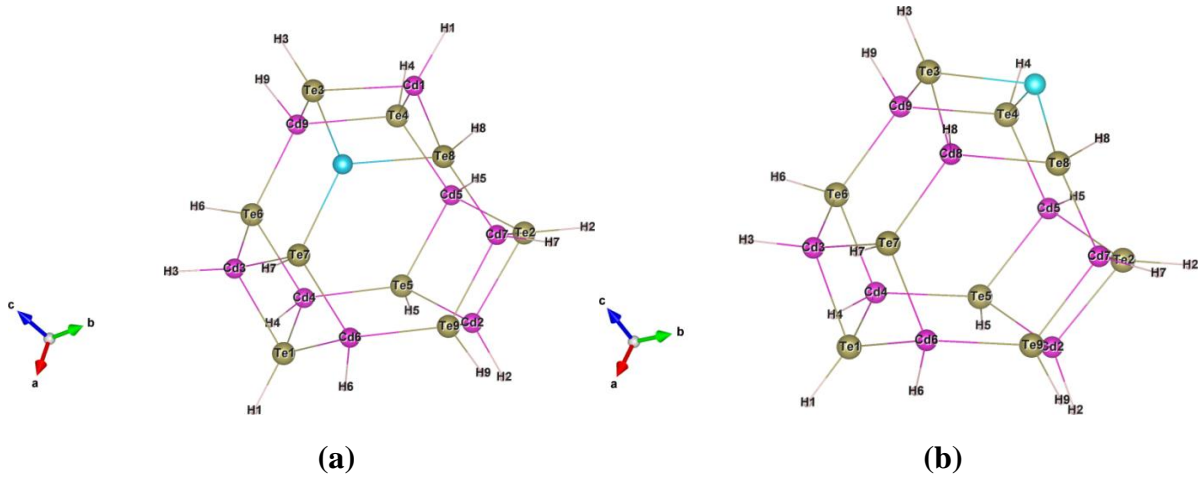

(a)  (b)

Figure S10. TM atom substituted for a Cd atom (a) in OR and (b) in CR of FHPA passivated $Cd_9Te_9$ cluster. The average TM-Te bond-lengths (nearest to TM i.e., Te7, Te8, Te3 for (a) and Te8, Te4, Te3 for (b)) are denoted as A and the bond-angles between those Te atoms are denoted as α. For each of those nearest Te as centre, like TM, similar average bond-lengths are represented as B, C, D and the average bond-angles as β, γ, δ respectively. In addition to this, the average angle between the TM and FHPA on nearest Te atoms are denoted as λ. X is the average FHPA-Te bond-length.

TM atom(s) substituted for Cd in CR in passivated cluster cause significant changes in bond-lengths and bond-angles. The average bond-angle Te-Cd-Te in undoped cases is 102.06 degree but when this Cd at CR is replaced by TM, the angle increases. This means that the TM atom moves towards the cluster interior. The reduced values of angles β, γ and δ with respect to undoped case show that the Te atoms move outwards. The average bond-length of Te-FHPA is increased for this case. This change is solely due to FHPA attached to a Te (Te3) in central ring that moves towards the TM atom resulting in an increase in Te-FHPA (Te3-H3) bond-length.

Table ST3. Contribution (acceptor, donor, deep levels) of the doped TM atom(s) in the energy gap of host cluster for different cases.

|  | Without passivation | | With passivation | |
|---|---|---|---|---|
|  | OR | CR | OR | CR |
| **Hole doping** | Cr, Co, Ni, Cu, Ru, Rh, Pd | Cr, Co, Ni, Cu, Rh, Pd | Mn, Fe, Co, Ni, Cu, Ru, Rh, Pd | Cr, Fe, Co, Ni, Cu, Rh, Pd |
| **Electron doping** | Ti, Mn, Fe | Ti, Mn, Fe | Ti | Ti, V, Zn, Ru |
| **Deep levels doping** | Ti, V, Cr, Fe, Co, Ni, Cu, Ru, Rh, Pd | Ti, V, Cr, Fe, Co, Ni, Cu, Ru, Rh, Pd | Ti, V, Cr, Mn, Fe, Co, Cu, Ru, Rh, Pd | Ti, V, Cr, Mn, Fe, Co, Ni, Cu, Zn, Ru, Rh, Pd |